\documentclass[%
 aip,
 amsmath,amssymb,
 reprint,%
]{revtex4-2}

\usepackage{graphicx}% Include figure files
\usepackage{dcolumn}% Align table columns on decimal point
\usepackage{bm}% bold math
\usepackage[utf8]{inputenc}
\usepackage[T1]{fontenc}
\usepackage{mathptmx}
\usepackage{xcolor}

\begin{document}

\preprint{AIP/123-QED}

\title[WinSPM]{{\color{blue}Simplified feedback control system for Scanning Tunneling Microscopy}}
% Force line breaks with \\

\author{Francisco Mart\'in-Vega}
\author{V\'ictor Barrena}
\author{Raquel S\'anchez-Barquilla}
\author{Marta Fern\'andez-Lomana}
\author{Jos\'e Benito Llorens}
\author{Beilun Wu}
\author{Ant\'on Fente}
\author{David Perconte Duplain}
\affiliation{
Laboratorio de Bajas Temperaturas y Altos Campos Magn\'eticos, Unidad Asociada (UAM/CSIC), Departamento de F\'isica de la Materia Condensada, Instituto Nicol\'as Cabrera and Condensed Matter Physics Center (IFIMAC), Universidad Aut\'onoma de Madrid, E-28049 Madrid,
Spain
}
\author{Ignacio Horcas}
\affiliation{ 
Departamento de F\'isica de la Materia Condensada, Universidad Aut\'onoma de Madrid, E-28049 Madrid,
Spain
}
\author{Raquel L\'opez}
\author{Javier Blanco}
\author{Juan Antonio Higuera}
\affiliation{ 
SEGAINVEX, Universidad Aut\'onoma de Madrid, E-28049 Madrid,
Spain
}
\author{Samuel Ma\~{n}as-Valero}
\affiliation{
Instituto de Ciencia Molecular (ICMol), Universidad de Valencia,
Catedr\'atico Jos\'e Beltr\'an 2, 46980 Paterna, Spain
}
\author{Na Hyun Jo}
\author{Juan Schmidt}
\author{Paul C. Canfield}
\affiliation{Ames Laboratory, U.S. DOE, Iowa State University, Ames, Iowa 50011, USA
}
\author{Gabino Rubio-Bollinger}
\affiliation{ 
Departamento de F\'isica de la Materia Condensada, Instituto Nicol\'as Cabrera and Condensed Matter Physics Center (IFIMAC), Universidad Aut\'onoma de Madrid, E-28049 Madrid,
Spain
}
\author{Jos\'e Gabriel Rodrigo}
\author{Edwin Herrera}
\author{Isabel Guillam\'on}
\author{Hermann Suderow}
 \email{hermann.suderow@uam.es.}
\affiliation{ 
Laboratorio de Bajas Temperaturas y Altos Campos Magn\'eticos, Unidad Asociada (UAM/CSIC), Departamento de F\'isica de la Materia Condensada, Instituto Nicol\'as Cabrera and Condensed Matter Physics Center (IFIMAC), Universidad Aut\'onoma de Madrid, E-28049 Madrid,
Spain
}

\date{\today}% It is always \today, today,
             %  but any date may be explicitly specified

\begin{abstract}
A Scanning Tunneling Microscope (STM) is one of the most important scanning probe tools available to study and manipulate matter at the nanoscale. In a STM, a tip is scanned on top of a surface with a separation of a few \AA. Often, the tunneling current between tip and sample is maintained constant by modifying the distance between the tip apex and the surface through a feedback mechanism acting on a piezoelectric transducer. This produces very detailed images of the electronic properties of the surface. The feedback mechanism is nearly always made using a digital processing circuit separate from the user computer. Here we discuss another approach, using a computer and data acquisition through the USB port. We find that it allows succesful ultra low noise studies of surfaces at cryogenic temperatures. We show results on different compounds, a type II Weyl semimetal (WTe$_2$), a quasi two-dimensional dichalcogenide superconductor (2H-NbSe$_2$), a magnetic Weyl semimetal (Co$_3$Sn$_2$S$_2$) and an iron pnictide superconductor (FeSe).
\end{abstract}

\maketitle

\section{Introduction}

G. Binnig and H. Rohrer invented the Scanning Tunneling Microscope (STM) back in 1981 \cite{Binnig1982}. The technique was soon extended to other probes based on a tip scanning a sample surface, revolutionizing the study and manipulation of materials at the nanoscale. The STM uses the quantum tunneling effect between an atomically sharp tip and a flat conducting sample through a vacuum barrier to sense the surface properties and the piezoelectric effect that provides precise subnanometric positioning of the tip over the sample. Usually, we measure the tunneling current between tip and sample and maintain its value constant using a feedback loop that acts on the piezo that controls the z-position of the tip. When scanning the tip in the x-y plane, the feedback signal provides the z-position of the tip. Thus, we can build two-dimensional (2D) maps of the surface at a constant tunneling current.

\begin{figure}[hbt]
\begin{center}
\includegraphics[clip=true,width=\columnwidth,keepaspectratio]{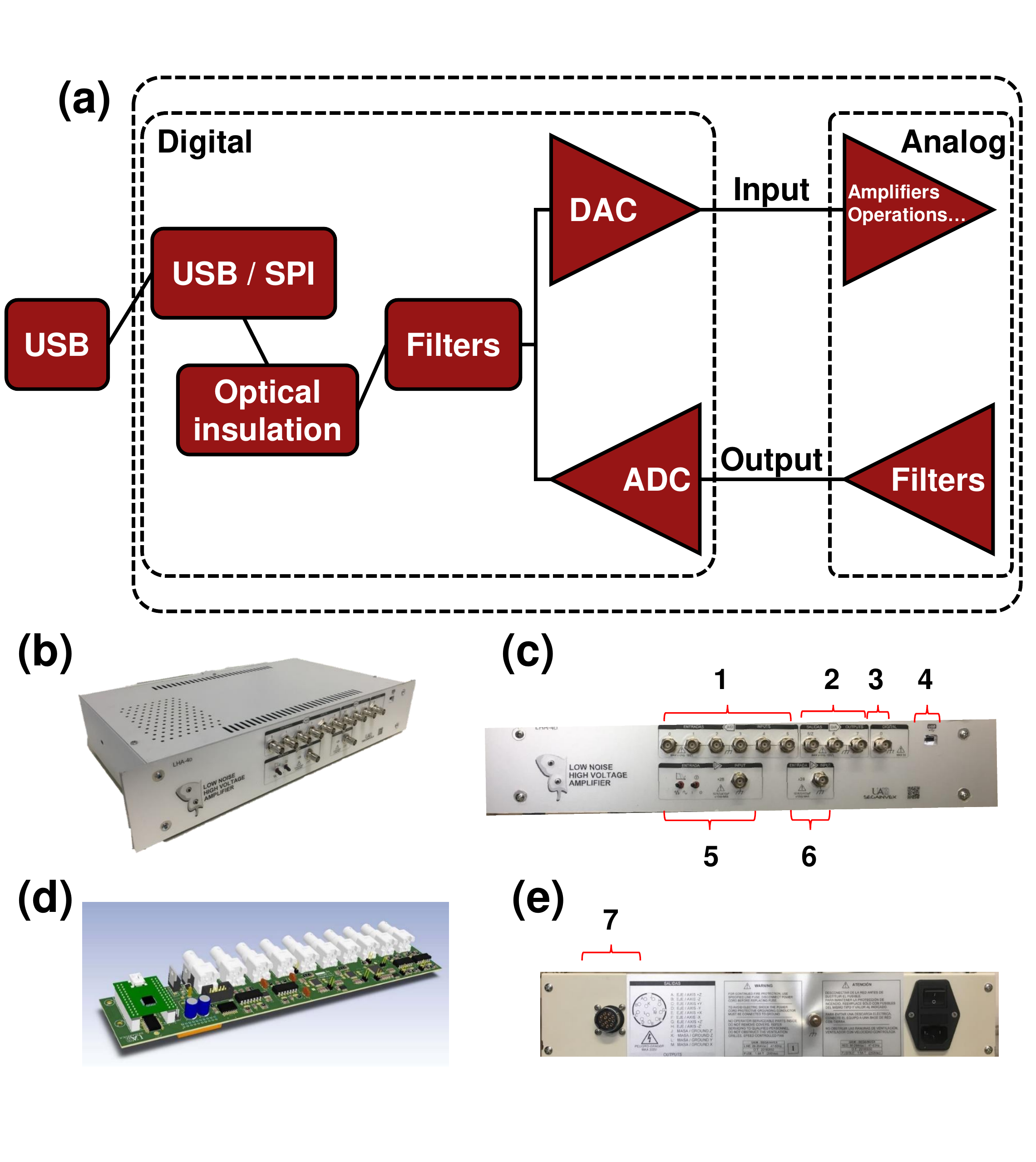}
\end{center}
\caption{(a) Block diagram of the digital and analog part of the STM electronics. The components are marked by red boxes. Black dashed lines mark the different parts. Digital and analog parts are in a single enclosure. (b) Picture of the device. (c) Front end of the device. The USB digital interface (d) is screwed to the front panel. There are five ADC channels for reading signals (1), three auxiliary DAC signals (2), one digital auxiliary output (3), the USB port entry (4), an input for the amplifier used for coarse motion (5) and an auxiliary input for the Z control (6). All high voltage outputs (7) are on the back side (e).}
\label{fig:Hardware}
\end{figure}

There are many different hardware designs to control scanning probe microscopes\cite{doi:10.1063/1.1140457,wiesendanger_1994,Voigtlander,Song2010,RevModPhys.75.949}. All STM designs are based on a kernel that includes a feedback mechanism which keeps a constant tunneling current during the scan. The feedback can be either analog\cite{Grafstrom_1990,Olin_94,FLAXER2006303,doi:10.1063/1.1147077,doi:10.1063/1.1149224,doi:10.1063/1.1145289,doi:10.1063/1.1141156} or digital\cite{doi:10.1063/1.1140589,doi:10.1063/1.5001312,doi:10.1116/1.3374719,doi:10.1063/1.1144462,Horcas2007}. It has been long thought that the digital feedback needs a real time acquisition and control system, using a digital signal processor (DSP) or a field programmable gate array (FPGA) that operates separately from the user computer. This indeed allows using a clock and taking data at fixed time intervals. However, STM often requires measurement of a parameter as a function of the position, not as a function of time. Here we show that one can successfully operate a STM using a simple USB based data acquisition on a computer running the Windows\textregistered\, operating system. This eliminates the need for defining fixed time intervals for data acquisition. We also show that our set-up allows for very low noise measurements. Addressing this problem usually requires careful filtering of the high frequency noise created at the control electronics\cite{Suderow2011,Song2010,Assig2013,Battisti2018,Misra2013,Machida2018,Marz2010,Li2012,doi:10.1063/1.4905531}.

Let us remind that J. Tersoff and D.R. Hamann applied Bardeen's tunneling theory to STM\cite{Bardeen1961,Tersoff1983,Tersoff1985} and showed that, under several simplifying assumptions and at zero temperature, the tunneling current vs.\,bias voltage $I(V_{Bias})$ is given by

\begin{equation}
I (V_{Bias}) \propto e^{-\frac{z}{z_{\phi}}}\int_{0}^{V_{Bias}}N_s(E-eV)N_t(E)dE
\end{equation}

where $z$ is the distance between tip and sample, $z_{\phi}\equiv \frac{\hbar}{\sqrt{8m\phi}}$, with $\phi$ the average workfunction,  $m$ the electron effective mass and $N_s(E-eV)$, $N_t(E)$ the energy dependent densities of states of tip and sample, respectively ($E$ being the energy). The tip is atomically sharp and its $N_t(E)$ does not depend on the position of the tip over the sample. Thus, the 2D maps of the surface at a constant tunneling current provide maps of the local density of states $N_s(x,y)$. The tunneling conductance $\frac{dI}{dV}$ as a function of the bias voltage is often propotional to the energy dependent $N_s(E=eV,x,y)$. $\frac{dI}{dV}$ can be recorded as a function of the position when scanning the tip on top of the surface. This leads to a set of 2D images, providing $N_s(E=eV,x,y)$. Controlling the noise level is particularly important when there is a need to resolve sharp features in the tunneling conductance, $\frac{dI}{dV}$, caused by variations of  $N_s(E)$ in a small energy range. This is often the case in cryogenic STM set-ups used to study materials with large variations in the electronic bandstructure at low energy scales. 

The simplified control electronics described here helps achieving the goal of low noise measurements more easily. We show results obtained in relevant materials and in several different set-ups which use the device described here. Software and full design of the electronics are available in, respectively Ref.\,\onlinecite{Drawings} and Ref.\,\cite{Code}. 

\section{Description of the set-up}

\subsection{Digital control system and computer interface}

We describe the hardware of our system in Fig.\,\ref{fig:Hardware}. The system has two 16 bits Digital to Analog (DAC) DAC8734 chips, each with four analog output lines, which provide voltage signals from $-10$ V to $+10$ V. These signals can be amplified by a factor of 14 using the high precision and ultra low noise amplifiers described in Ref.\,\onlinecite{doi:10.1063/1.4905531}. Acquisition of analog voltage signals is done using a 16 bit Analog to Digital (ADC) chip (ADC7606) with eight separate input channels.

\begin{figure}[hbt]
\begin{center}
\includegraphics[clip=true,width=\columnwidth,keepaspectratio]{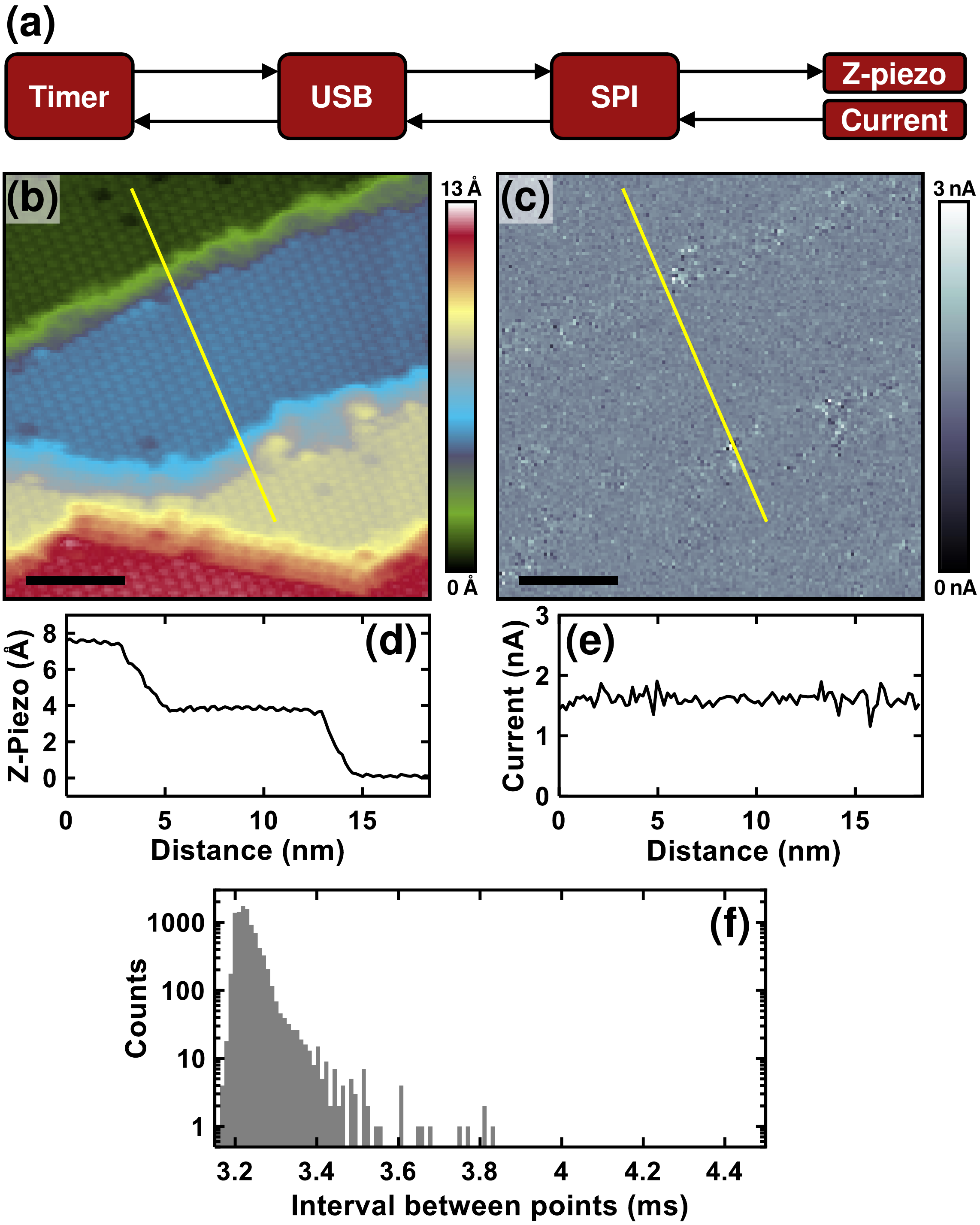}
\end{center}
\caption{(a) Block diagram of the controlling system. (b-e) Example of a scanning experiment where the feedback loop is functioning to maintain a constant current, with the computer running as usual. In (b) we show the feedback signal as a function of the position, converted to the z-position of the tip in \AA. We can see terraces and atomic resolution, obtained on Co$_3$Sn$_2$S$_2$ scanned with a tip of Au (temperature of 4.2 K, magnetic field of 14 T, bias voltage of 100 mV and current of 1.6 nA). The height color scale is given on the right. In (c) we show the current taken at each position of the image in (b). In (d) we show the height along one scan line and in (e) the current read along the same line (yellow line in (b,c)). Black scale bars in (b) and (c) are 5 nm long. (f) Interval between two points when measuring a signal as a function of time during a 30 second period approximately each 3.2 ms.}
\label{fig:PhotoElectronics}
\end{figure}

\begin{figure}[hbt]
\begin{center}
\includegraphics[clip=true,width=\columnwidth,keepaspectratio]{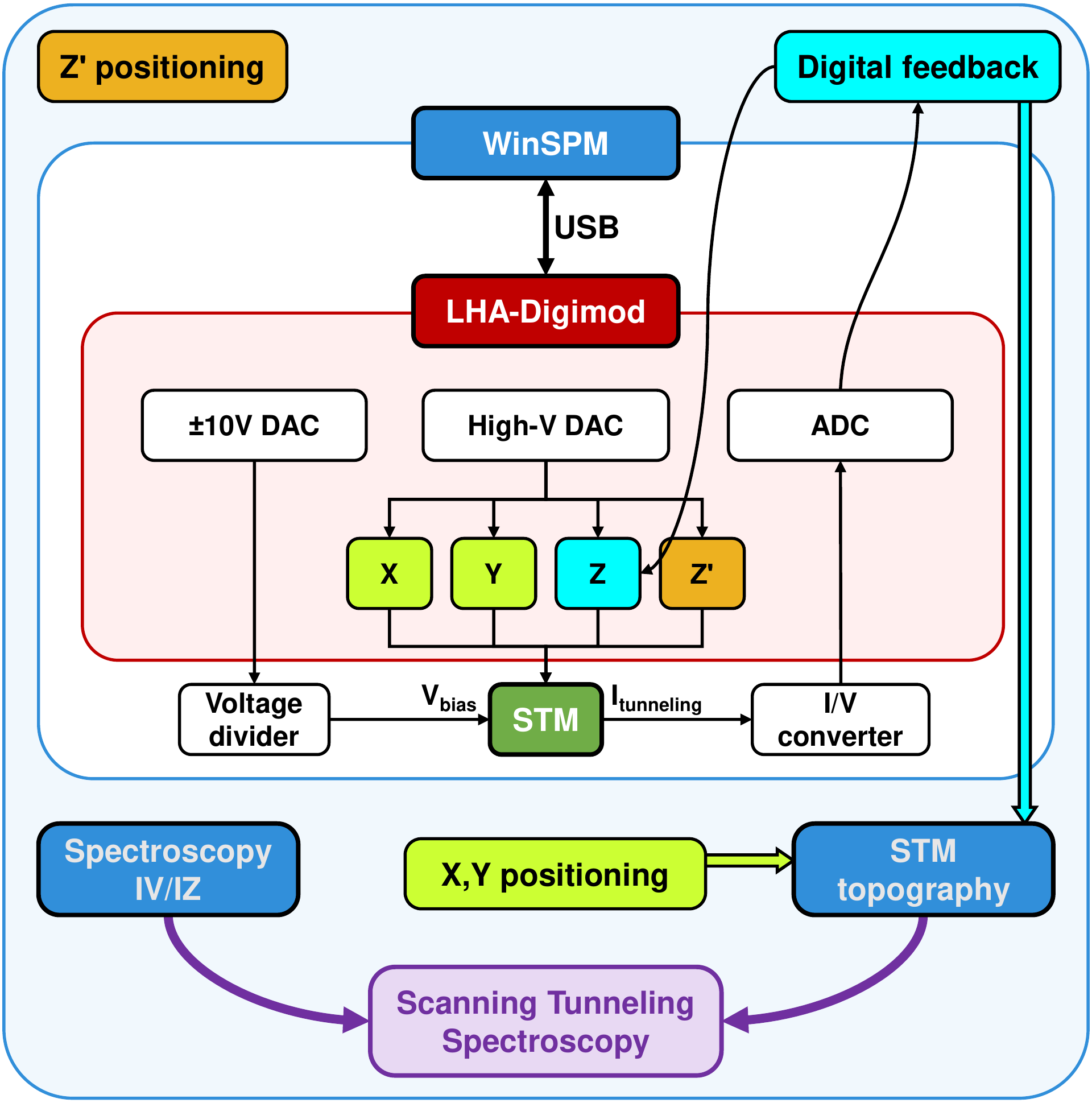}
\end{center}
\vspace*{-0.5cm}
\caption{Block diagram of the different operations of the data acquisition software and its communication with the digital electronics. There is a coarse positioning module ("Z' positioning"), a spectroscopy module to take current vs.\,voltage IV, or current vs.\,distance IZ curves ("Spectroscopy IV/IZ") or other curves (such as the conductance vs bias voltage) and an image module ("STM topography"). The program controls the electronics (red box) through a USB controlled interface called LHA-Digimod which is located on the front of the electronics (Fig.\,\ref{fig:Hardware}(c)). The STM set-up, with the piezotube, approach system, tip and sample, is indicated by the dark green box marked "STM". Some additional equipment as the voltage divider and the current to voltage converter are indicated in white boxes. The digital feedback module, schematically shown in light blue, runs to maintain a constant current during operation, fed by the current measurement and acting on the Z-piezo. The signal sent to the Z-piezo is used to trace the STM topography.}
\label{fig:diagram_acquisition}
\end{figure}

To address the DAC and ADC system, we use a USB 2.0 FIFO (first in first out) circuit (FTDI2232H) which addresses the serial peripheral interface (SPI). The USB is optocoupled to the board to avoid interference from the main computer. All SPI signals are filtered using one pole RC filters with a cutoff frequency slightly above the frequency of the clock required to address the DAC and ADC with the SPI (2 MHz). Outputs are carefully low pass filtered using analog one pole RC filters cutting at 10kHz (Fig.\,\ref{fig:Hardware}(a)). The digital part of the electronics (Fig.\,\ref{fig:Hardware}(d)) is mounted on a multiple layer board, taking care of separating as far as possible lines where the SPI bus runs from the signal in and output lines. Furthermore, it has a separate power supply and is mounted very close to the high voltage amplifiers to avoid picking up noise. Grounding is designed to avoid induction from digital signal lines by minimizing the impedance with large ground plates and a full connection to the metallic enclosure. In addition, microtransformers separate communication bus from signals to avoid interference.  The noise of the output of the DAC is below 1$\mu$V RMS and the ADC has an analog filter and a digital sampling averaging system, giving a high signal to noise ratio. The wiring between the electronics and the microscope is made by carefully mixing cables that hold the signals with ground cables, firmly attached to the electronics and to the cryostat.

Five of the amplified output voltage signals are used to drive the scanner piezotube, four for X and Y motion and one for the Z motion. X and Y motion is performed using two DACs mounted in cascade, achieving thus $\mu$V resolution. This set-up provides enough high voltage lines to operate a usual cryogenic microscope, as the ones discussed in Refs.\cite{Suderow2011,Song2010,Assig2013,Battisti2018,Misra2013,Machida2018,Marz2010,Li2012,doi:10.1063/1.4905531}. Another amplified voltage output line is used to drive the coarse approach motor. There are three further low voltage lines for the bias voltage and additional needs.

The electronics hardware is handy (Fig.\,\ref{fig:Hardware}(b)) and the whole system can be easily transported and connected to the USB port of a computer. Different systems can be used on the same computer, by using separate USB ports.

\subsection{Description of the control software}

\begin{figure}
\begin{center}
\includegraphics[clip=true,width=0.79\columnwidth,keepaspectratio]{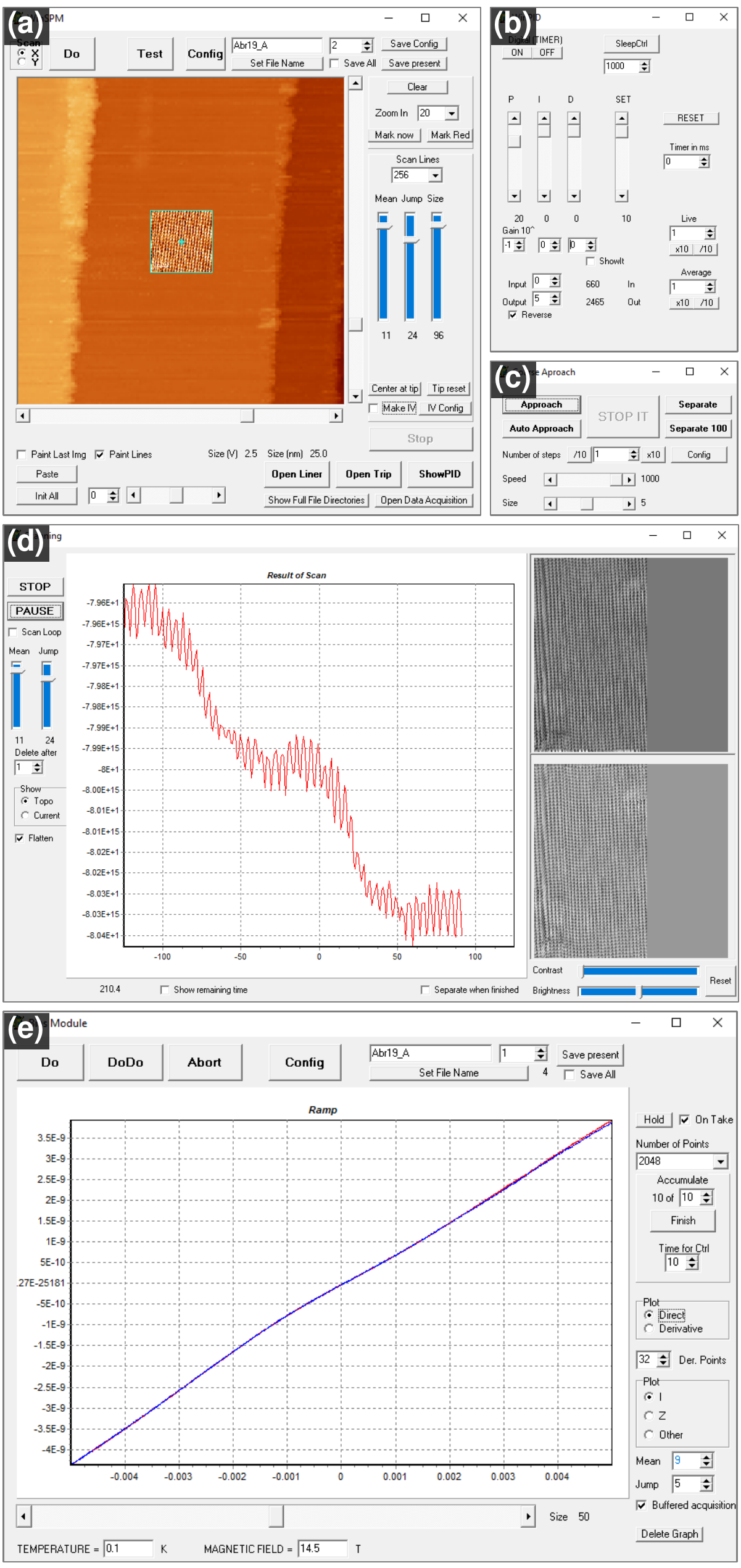}
\end{center}
\caption{Screenshots from the data acquisition software. (a) Scanner interface. Blue cross and blue square mark the position of the tip and the scanning area, respectively. We depict terraces and an atomically resolved image on one terrace as colored images (from black to yellow to white). Data are obtained on FeSe at 100 mK and at a magnetic field of 14.5 T with a bias voltage of 5 mV and 4 nA tunneling current. (b) Digital PI (proportional integral) feedback interface window. (c) Coarse vertical motion (Z$^{‘}$) control window. (d) Interface for the live preview of the scan. The central panel shows the live tip height profile and the progress of the topographic images made by scanning on the $\pm$ Y directions is shown in the right panels (top panel for scanning along one direction in Y and bottom panel for scanning back). (e) Curve acquisition interface, showing the current vs.\,voltage as blue and red lines.}
\label{fig:screenshots_acquisition}
\end{figure}

We have written communication routines in Delphi\cite{Code}. The feedback is made by reading the current and sending an output signal calculated using a PI (proportional integral) algorithm (Fig.\,\ref{fig:PhotoElectronics}(a)). The latency of USB port allows to access the port with a bandwidth reaching a few kHz. This is below the typical resonance frequency of the STM, as required to avoid noise amplification and oscillations\cite{Voigtlander}. We show an example of an obtained image on Co$_3$Sn$_2$S$_2$ using a Au tip in Fig.\,\ref{fig:PhotoElectronics}(b-e). Further examples of similarly acquired images in other compounds and under different conditions are given below.

To have an optimized time base for execution we use threaded timer and program its access in such a way that it is regularly called during time consuming operations such as loops for data saving\cite{timers}. This provides smooth operation without interruptions\cite{RealTime2,RealTime3}. We have followed data acquisition and PI operation for many hours, acquiring billions of points and observing random interruptions of the latency which were significantly below 1 ms and just a few interruptions requiring up to 10 ms.  None of these interruptions lead to a tip instability, as seen for example in Fig.\,\ref{fig:PhotoElectronics}(b-e), which provides raw data (no corrected points). To better quantify the jitter in the latency, we have measured a signal as a function of time with a wanted time interval of 3.2 ms. In Fig.\,\ref{fig:PhotoElectronics}(f) we show the interval between two measurements. We see that we have a deviation of 0.2 ms from the wanted interval in less than 1\% of acquired points and deviations larger than 0.4 ms in far less than 0.1\% of the acquired points. Thus, although the system is not a priori thought for real time measurements, it can be used in measurements as a function of time where a constant time interval is not critical in the acquisition.
 
We show a block diagram of the components of the software (blue background) and how these are linked to the hardware (red and white backgrounds) in Fig.\,\ref{fig:diagram_acquisition}. The software interacts with the hardware through a main window ("WinSPM"). The main window includes two additional windows. One is used to make spectroscopy ("Spectroscopy IV/IZ") or any other operation that requires cutting the feedback loop. And the other one ("STM topography") is used to show and control the scanning process. The feedback ("Digital feedback") operates continously in the background, reading the current value and acting on Z electrode of the piezotube through the corresponding amplifier, whose value is also sent to "STM topography" to monitor the Z as a function of the position of the tip during scanning. The main window handles the USB communication through the interface containing the DAC and ADC converters ("LHA Digimod"). One converter is used for setting the bias voltage and others are amplified to set the x, y and z positions. The ADC is used, for instance, to read the current from the current to voltage transimpedance amplifier ("I/V converter"). In addition, another high voltage amplifier is used to drive the coarse approach motor. During imaging, we can stop at each point and perform IV or bias voltage vs conductance curves, to obtain images of the current or the conductance as a function of the bias voltage ("Scanning Tunneling Spectroscopy").

\begin{figure}[hbt]
\begin{center}
\includegraphics[clip=true,width=\columnwidth,keepaspectratio]{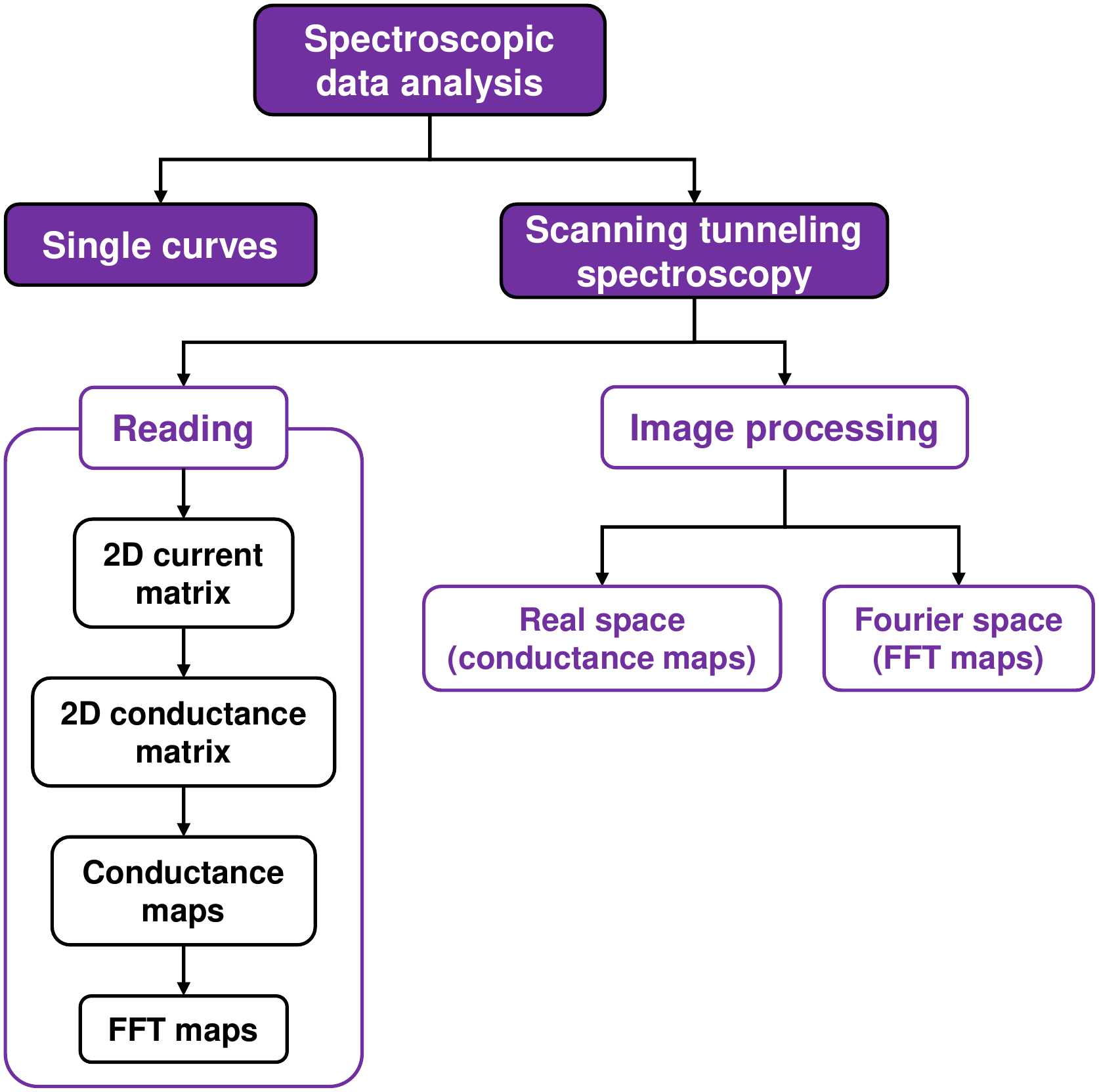}
\end{center}
\caption{Block diagram of the data analysis software. The data are stored in a binary file with the required headers to return image size, position, etc. There is a curve reading module ("single curves"), which allows rendering the data and performing basic operations to understand tunneling current curves, such as derivation, temperature deconvolution and plots including many curves. The "scanning tunneling spectroscopy" module is separated into a "reading" module which creates current and conductance matrices, and from these, conductance and Fourier transform maps at given bias voltage values. The "image processing" module allows rendering the images in a user friendly manner and making different operations required to understand real and Fourier space images. The latter is easily extended using Matlab\textregistered\, functions.}
\label{fig:diagram_analysis}
\end{figure}

We provide an example of screenshots during a typical experiment in Fig.\,\ref{fig:screenshots_acquisition}. The experiment is made at 100 mK, with a tip of Au and a sample of FeSe. In Fig.\,\ref{fig:screenshots_acquisition}(a) we show the main working window, where we have pasted an image showing steps in FeSe. The large square shows an image of an area of several nm square, showing steps in FeSe, and another image taken around the central area showing atomic resolution. The blue cross in the center of the square is the position of the STM tip. In Fig.\,\ref{fig:screenshots_acquisition}(d) we show a typical scan (red line in the left panel) and the image being build up on the right panels. We can clearly see  the profiles of Se atoms in the line scan and the square atomic Se lattice in the images. In Fig.\,\ref{fig:screenshots_acquisition}(e) we show a current vs.\,voltage curve (blue and red overlapped lines) obtained in FeSe under a magnetic field of 14.5 T. We see that the curve is slightly non-linear due to the superconducting gap opening.

To obtain maps of IV or tunneling conductance vs bias voltage curves at each point of an image, we disconnect the feedback loop and perform the measurement at each point of the scan. We store separately the topography file with usual scanning information and the IV curves.

\begin{figure*}[hbt]
\begin{center}
\includegraphics[clip=true,width=\textwidth,keepaspectratio]{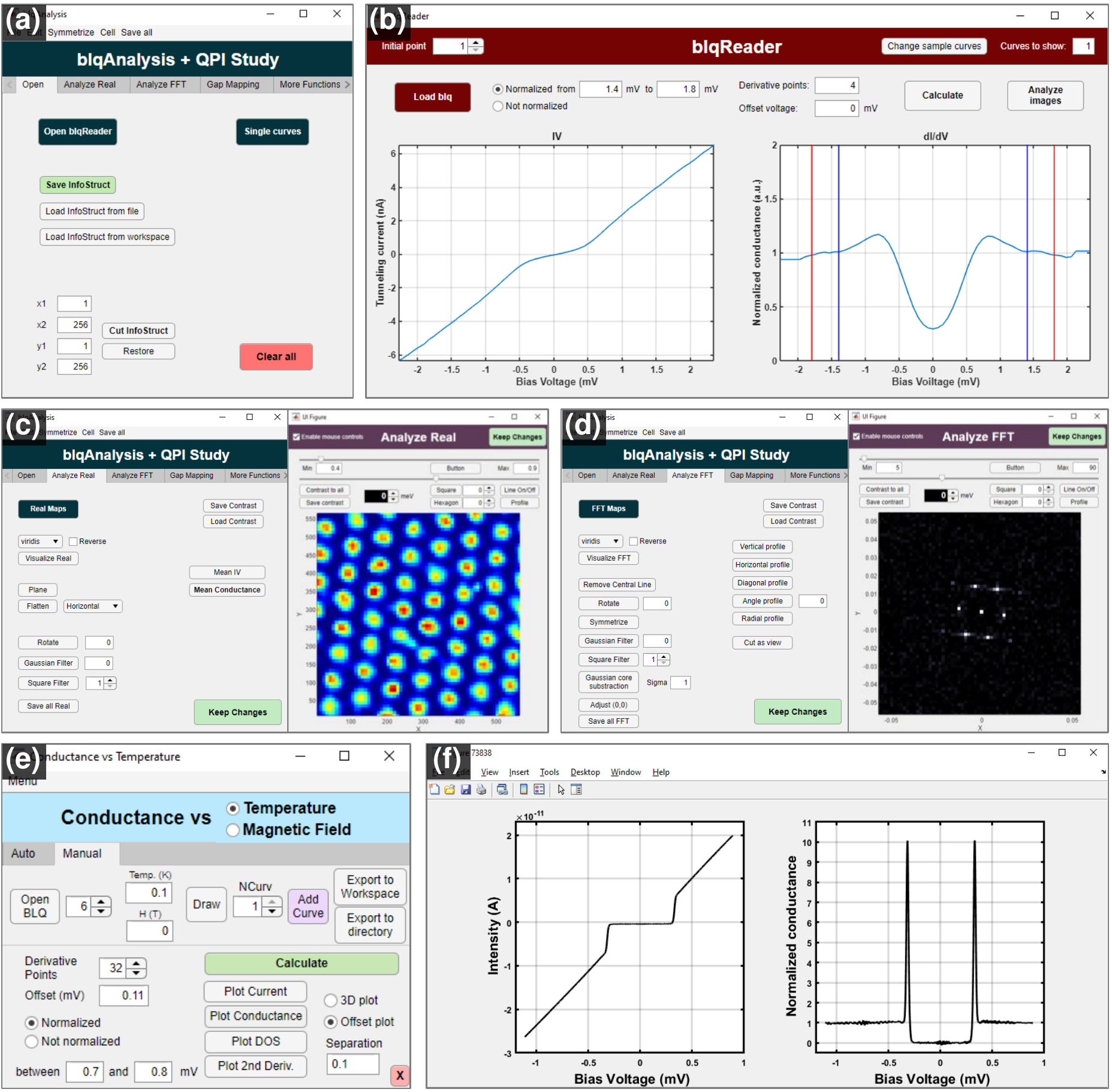}
\end{center}
\caption{Screenshots from the data analysis software. (a) Main window of the program, with the different modules separated in menu items. (b) Conductance curves calculation window. We can see the superconducting gap at a magnetic field of 0.1 T obtained in $\beta-$Bi$_2$Pd as blue lines. (c) Window for viewing conductance maps. We show a picture of the vortex lattice in $\beta-$Bi$_2$Pd  obtained from the zero bias conductance map. (d) Window showing the Fourier transform of the image in (c). A number of operations can be performed on the Fourier transform, like rotating, filtering, symmetrizing, etc. (e) Main window to plot and make calculations on tunneling current and tunneling conductance vs bias voltage curves. (f) Tunneling current (left panel) and tunneling conductance (right panel) obtained using Al tip and sample below 100 mK.}
\label{fig:screenshots_analysis}
\end{figure*}

\subsection{Description of the analysis software}

The file containing all IV (or tunneling conductance vs bias voltage) curves is actually a 3D matrix with an IV (or tunneling conductance) curve at each pixel of a 2D space that spans the real space image. We analyze these files using a software based on the Matlab\textregistered\, environment \cite{MATLAB}. Related software based on Matlab\textregistered\, can be found in Ref.\cite{OtherCode}. Our software includes numerous features required to analyze STM images, such as Fourier transform spectroscopy, rotation and manipulation of images, plot of conductance maps at any bias voltage and obtaining and analyzing tunneling conductance curves from anywhere in the images.

We show a block diagram of the data treatment software in Fig.\,\ref{fig:diagram_analysis}. This includes a module for single curves, a reading module and an image processing module.

The reading module needs two main inputs: the $n$ by $m$ dimensions of the topographic image, the corresponding binary file containing the associated spectroscopic curves and the starting point in the binary file. We then average data using adjacent points around an interval $\delta(V)$. Normalization with respect to the conductance value at a certain bias voltage range can be used to eliminate setpoint effects, as discussed in Ref.\,\onlinecite{Lawler2010}. Data can be saved into a structure object that stores all the variables needed for a further image processing analysis.

\begin{figure}[hbt]
\begin{center}
\includegraphics[clip=true,width=\columnwidth,keepaspectratio]{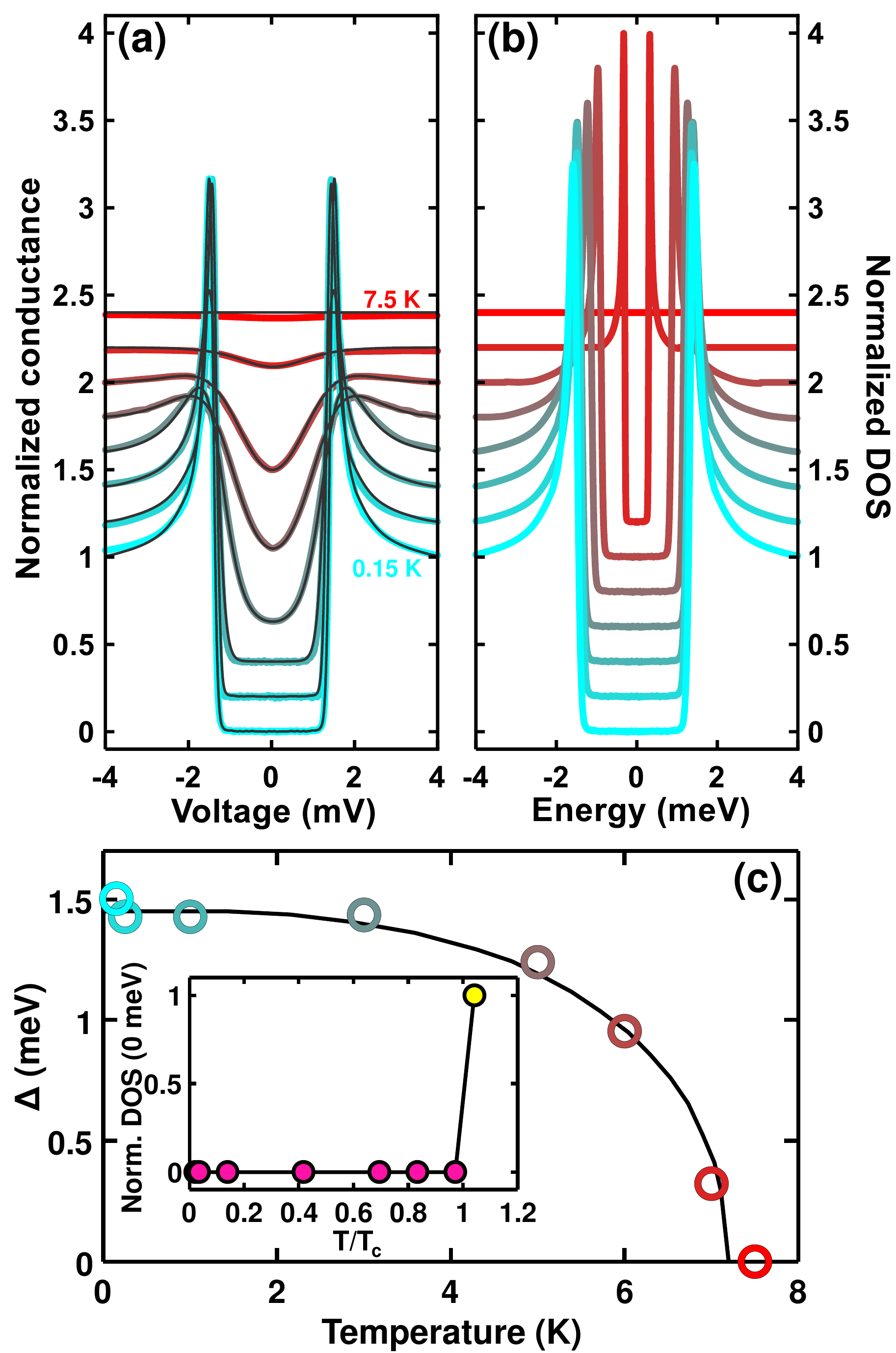}
\end{center}
\caption{(a) Tunneling conductance vs.\,bias voltage taken at 0.15 K, 0.25 K, 1 K, 3 K, 5 K, 6 K, 7 K and 7.5 K in Pb, using a tip of Au. The black lines are the convolution of the density of states (shown in (b)) with the Fermi function. The position of the quasiparticle peaks provides the value of the superconducting gap, shown in the main panel of (c). Black line is a fit to BCS temperature dependence of the superconducting gap $\Delta(T)$. In the inset we show the density of states at zero energy as a function of temperature. We see that it remains zero up to the point when superconductivity vanishes.}
\label{fig:deconvolution}
\end{figure}

We show an example and screenshots in Fig.\,\ref{fig:screenshots_analysis}. In Fig.\,\ref{fig:screenshots_analysis}(a) we show the acquisition module, which we use to open files and set a few relevant parameters. In Fig.\,\ref{fig:screenshots_analysis}(b) we show the IV curve reader, with an IV curve taken in $\beta-$Bi$_2$Pd at 100 mK and at 0.1 T shown as a blue line. We show the IV curve (left panel) and the tunneling conductance vs.\,bias voltage (right panel). The red and blue vertical lines in the right panel show the voltage range which we use to normalize curves in this particular case. The resulting maps of the tunneling conductance are shown in Fig.\,\ref{fig:screenshots_analysis}(c,d), where we plot the zero bias conductance as a function of the position (Fig.\,\ref{fig:screenshots_analysis}(c)) and its Fourier transform (Fig.\,\ref{fig:screenshots_analysis}(d)). The images show the superconducting vortex lattice of  $\beta-$Bi$_2$Pd at 0.1 T and 100 mK\cite{RevModPhys.79.353,Suderow2014,Herrera2015}. We can visualize results at different bias voltages, identify vortices or triangulate vortex lattices \cite{Guillamon2009,Guillamon2014,PhysRevResearch.2.033133}. We can also use the same modules to visualize oscillations in the tunneling conductance due to impurity scattering, called quasiparticle interference. Quasiparticle interference measures the electronic bandstructure\cite{Hoffman1148,Sprunger1764,McElroy2003,PhysRevLett.107.186805,Hoffman2011,Lee2009,Iwaya2017,Xu2015,Inoue2016,PhysRevB.103.L060506}. We can perform different operations of quasiparticle interference, as rotations, symmetrization, filtering or extraction of profiles. The module for single curves is shown in Fig.\,\ref{fig:screenshots_analysis}(d), with an example of an IV curve and the corresponding conductance obtained by using Al as tip and sample.

\begin{figure}[hbt]
\begin{center}
\includegraphics[clip=true,width=\columnwidth,keepaspectratio]{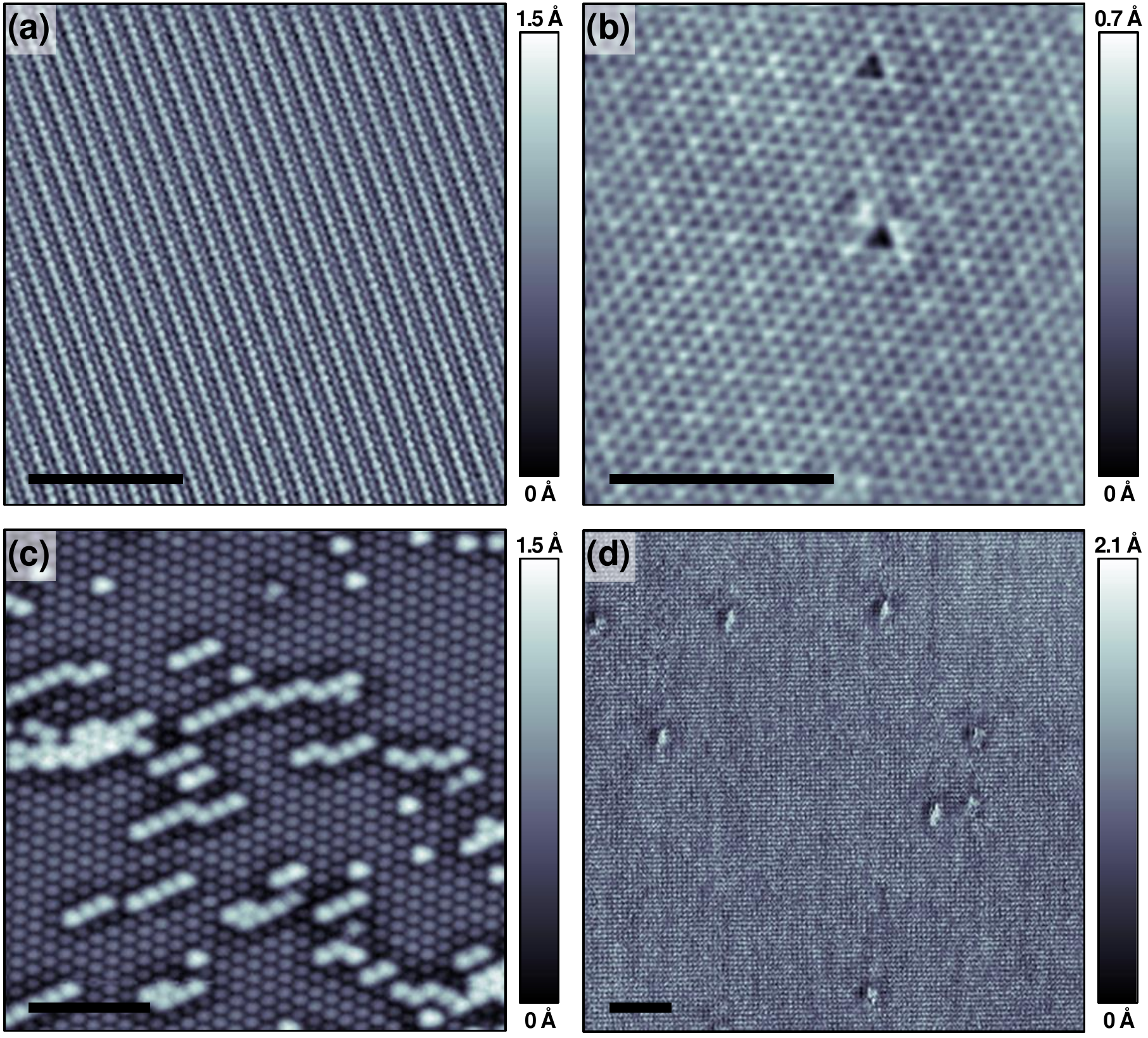}
\end{center}
\caption{(a) Atomic resolution image obtained in WTe$_2$, with a bias voltage $V_{bias}=100$ mV and tunneling current $I_{tun}=4$ nA. (b) STM image obtained in 2H-NbSe$_2$ with $V_{bias}=100$ mV and $I_{tun}=3.2$ nA. (c) STM image obtained in Co$_3$Sn$_2$S$_2$ with $V_{bias}=100$ mV and $I_{tun}=1.6$ nA. (d) STM image obtained in FeSe, with $V_{bias}=10$ mV and $I_{tun}=4$ nA. All images have been taken at a temperature of 4.2 K (a,b,c) and 100 mK (d). Scale bars in black are 2 nm long. Vertical bars provide the z-changes as a color scale. Images are raw, except for the removal of a plane.}
\label{fig:examples}
\end{figure}

As an example of data treatment, we can discuss tunneling spectroscopy vs.\,temperature measurements in Pb, using a tip of Au, shown in Fig.\,\ref{fig:deconvolution}. The experimental curves obtained are shown as colored lines in Fig.\,\ref{fig:deconvolution}(a). To understand these curves, we construct a density of states as a function of the energy $N(E)$ (shown in Fig.\,\ref{fig:deconvolution}(b), bottom light blue line). $N(E)$ closely follows BCS expression ($N(E)=\frac{E}{\sqrt{E^2-\Delta^2}}$ with $\Delta$ the superconducting gap), with the value of the superconducting gap expected for Pb ($\Delta=$ 1.37 meV), except that the divergence at $eV=\Delta$ is substituted by non-infinite quasiparticle peaks. This can be due to a combination of inelastic non-equilibrium quasiparticles, the anisotropy of the superconducting gap of Pb and the finite energy resolution of our experiment\cite{PhysRevLett.114.157001,Rodrigo2004a,Suderow2011,Song2010,Assig2013,Battisti2018,Misra2013,Machida2018,Marz2010,Li2012,doi:10.1063/1.4905531}. Using this $N(E)$ we can then calculate the tunneling conductance for all temperatures, by convoluting with the derivative of the Fermi function. The results are shown as black lines in Fig.\,\ref{fig:deconvolution}(a). To obtain the tunneling conductance, we modify $N(E)$ at each temperature until we reproduce the experiment. From the position of the quasiparticle peaks in $N(E)$ we obtain $\Delta(T)$. We see that it follows the BCS superconducting gap dependence, $\Delta(T)$, shown in Fig.\,\ref{fig:deconvolution}(c).

\section{Results obtained in several STM cryogenic set-ups.}

 We now operate in our laboratory several units of the system described here. In Fig.\,\ref{fig:examples} we show results of topographic images at constant current in WTe$_2$, 2H-NbSe$_2$, Co$_3$Sn$_2$S$_2$ and FeSe. We see the one-dimensional Te chains that usually appear on the surface of WTe$_2$ (Fig.\,\ref{fig:examples}(a)). We also see the typical atomic lattice and charge density wave that is characteristic of 2H-NbSe$_2$ (Fig.\,\ref{fig:examples}(b)). The charge density wave is hexagonal and produces a charge modulation which is nearly commensurate with a period three times the lattice constant. There are two vacancies visible in the surface lattice, which consists of a Se hexagonal lattice. In Co$_3$Sn$_2$S$_2$ (Fig.\,\ref{fig:examples}(c)) we observe an atomically resolved surface, consisting of two different atomic species. The continuous hexagonal layer is composed of Sn atoms. Notice their hexagonal shape. The upper incomplete layer is made of S atoms. There is a tendency to form rows and the S atoms induce shadow-like features in the neighboring Sn atoms. In FeSe (Fig.\,\ref{fig:examples}(d)) we observe the square Se lattice. There are defects with an elongated in-plane shape, characteristic of the nematic features of the electronic properties of this compound.

The images are provided raw, implying that, during their acquisition which needed between half and hour and five hours, the data acquisition was fast enough so that there were no significant interruptions that influenced the stability of the PI algorithm controlling the tip z-position.

\section{Discussion and outlook}

In summary, we have described a system where we successfully control a STM microscope using a simple USB based data acquisition. Details of hardware and software are available in, respectively, Ref.\,\onlinecite{Drawings} and Ref.\,\onlinecite{Code}.

Our approach builds on developments of so-called soft real time data acquisition, as opposed to data acquisition on an exactly periodic time basis\cite{RealTime2,RealTime3}. Our results show that for STM and probably for many other applications too, computers allow for an access latency for data acquisition in usual in and output ports that is enough for succesful operation.

Care should be taken to use the method proposed here in combination with a PI algorithm. The system stability should be tested, so as to make sure that random interruptions do not influence the PI stability. At least for the systems imaged here, and probably for many other systems which require high resolution and thus slow data acquisition, the latency for accessing the port of a usual desktop computer is largely sufficient to maintain the stability of the PI. Furthermore, measurements as a function of time require obviously a stable time basis. This can be circumvented to some extend by measuring the time difference between acquisition, but leads to curves with random changes in the time base. Our results show that these changes are mostly below a ms and only rarely extend over more than a few ms.

There are commercially available choices for data acquisition using the USB port that include data processors\cite{commercialni,commercialredpit}. Using appropriate filtering, a similar noise level as we obtain here could be achieved\cite{doi:10.1063/1.5001312}. However, we believe that its implementation is more difficult than using the simple solution proposed here. Furthermore, calculations required to acquire the actual data are made within the data acquisition systems. We believe instead that allowing the user to perform operations on the main controlling computer provides a significant improvement. Multiple core computers with different components running in parallel are becoming largely available. This will produce a shift towards integrating data acquisition within the running operational system environments. The results obtained here could then be applied to all kinds of Scanning Probe Microscopes. The user will be able to considerably increase the possibilities for making complex operations during scanning.

\begin{acknowledgments}
We acknowledge discussions with Nicol\'as Agra\"it and Sebasti\'an Vieira. This work was supported by the Spanish Research State Agency (FIS2017-84330-R, CEX2018-000805-M, RYC-2014-15093, MAT2017-87134-C2-2-R, MAT2017-88693-R), by the Comunidad de Madrid through program NANOFRONTMAG-CM (S2013/MIT-2850) and by the European Research Council PNICTEYES through grant agreement 679080. We acknowledge collaborations through EU program Cost CA16218 (Nanocohybri). Work at Ames Laboratory was supported by the U.S. Department of Energy, Office of Basic Energy Science, Division of Materials Sciences and Engineering. Ames Laboratory is operated for the U.S. Department of Energy by Iowa State University under Contract No. DE-AC02-07CH11358. NHJ and JS were also supported by the Gordon and Betty Moore Foundation’s EPiQS Initiative through Grant GBMF4411.
\end{acknowledgments}

\textbf{Data availability.} Data available on request from the authors.

%\nocite{*}
%\bibliography{Win_SPM}% Produces the bibliography via BibTeX.

%aipnum4-2.bst 2019-01-14 (MD) hand-edited version of apsrev4-1.bst
%Control: key (0)
%Control: author (8) initials jnrlst
%Control: editor formatted (1) identically to author
%Control: production of article title (0) allowed
%Control: page (1) range
%Control: year (1) truncated
%Control: production of eprint (0) enabled
\providecommand{\noopsort}[1]{}\providecommand{\singleletter}[1]{#1}%

\end{document}